\documentclass{aa}  

\usepackage{graphicx}
\usepackage{txfonts}
\usepackage{xcolor}
\begin{document}

   \title{Testing volume integrals of travel-time sensitivity kernels for flows}

   \author{Michal \v{S}vanda
          \inst{1,2}
          \and
          Daniel Chm\'urny\inst{1}
          }

   \institute{Astronomical Institute, Charles University, Faculty of Mathematics nad Physics, V Hole\v{s}ovi\v{c}k\'ach 2, CZ-18000 Prague, Czech Republic\\
              \email{svanda@sirrah.troja.mff.cuni.cz}
         \and
             Astronomical Institute of the Czech Academy of Sciences, Fri\v{c}ova 298, CZ-25165 Ond\v{r}ejov, Czech Republic }

   \date{Received -; accepted -}

 
  \abstract
   {Helioseismic inversions rely largely on sensitivity kernels, 3-D spatial functions describing how do the changes in the solar interior translate into the changes of helioseismic observables. These sensitivity kernels come in most cases from the forward modelling utilising state-of-the-art solar models. }
   {We aim to test the sensitivity kernels by comparing their volume integrals with measured values from helioseismic travel times. }
   {By manipulating the tracking rate, we mimic the additional zonal velocity present in the Dopplergram datacubes. These datacubes are then processed by a standard travel-time measurements pipeline. We investigate the dependence of the east--west travel time averaged over a box around the disc centre on the implanted tracking velocity. The slope of this dependence is directly proportional to the total volume integral of the sensitivity kernel corresponding to the used travel-time geometry. }
   {We find a very good to acceptable agreement between measurements and models for travel times with ridge filtering applied. The sought dependence indeed resembles a linear function and the slope of it agrees with the expected volume integral from the forward-modelled sensitivity kernel. The agreement is less optimistic for the phase-speed filtered datacubes. A disagreement is particularly large for smallest phase speeds (filters td1--td4), for the larger phase speeds, our result indicate that the measured kernel integrals are systematically larger than expected from the forward modelling. We admit that for large phase-speeds and higher radial modes our testing procedure does not have to be appropriate.}
   {}


   \maketitle
%

\section{Introduction}
\subsection{Local helioseismology}
The photospheric layers of the Sun are optically thick, prohibiting us from directly studying the interior of our mother star. It turned out in the last seven decades that luckily, some of the information hiding under the impenetrable solar skin may be revealed indirectly, using the helioseismic method. 

Helioseismology \citep{2005LRSP....2....6G,2016LRSP...13....2B} in general encompasses a set of methods enabling us to examine the inner composition and dynamics of the Sun through the analysis of solar oscillations. There were many aspects of the solar interior structure that were successfully revealed thanks to the helioseismology. The depth of the convection zone, character of the rotation of the convective envelope, global meridional mass circulation or hints on pre-emergence of active regions to name a few. 

The helioseismology is based on observation and interpretation of the manifestations of seismic waves that propagate through the interior. These waves are excited near the solar surface likely by the vigorous convection within solar granules. The travel path (in the first approximation) is determined by the 1-D structure of the Sun. Perturbations of this ``background'' structure cause the waves to deviate in propagation and change the speed \citep{christensen2008lecture}. The waves manifest themselves in some surface observables, such as continuum intensity, or Doppler shifts and line-depths of photospheric spectral lines and others. They are best seen in the above mentioned Doppler shifts. Hence, state-of-the-art helioseismic instruments record spatially resolved Doppler shifts of a particular photospheric spectral line -- the Dopplergrams. The sequence of the Dopplergrams recording dynamical changes at a particular spatial position are then processed in order to learn about the behaviour of the solar seismic waves. 

A specific approach termed the \emph{local helioseismology} \citep[see a review by][]{2010ARA&A..48..289G} focuses to observation and analysis of the waves in the spatially limited region, the wave field within the chosen region is influenced by the properties of that particular area, both at the surface and beneath it. Consequently, by analyzing the wave patterns in this local region, it becomes possible to deduce local 3-D structures and dynamics beneath the surface \citep{dalsgaard2002}. The standard procedure used in all methods of local helioseismology involves selecting a small region on the solar surface and tracking it within a co-rotation frame, i.e. the region does not move in time relative to the surface of the Sun. 

A specific method of local helioseismology, the \emph{time--distance} helioseismology \citep{duvall1993}, deals with the travel times of the waves that are tracked when travelling between two places on the Sun. In its simplicity, the travel time $\tau$ is measured between two measurement points (pixels) in the surface measurements,  $\mathbf{r}_1$ and  $\mathbf{r}_2$. Let us indicate helioseismic observable (e.g. a sequence of Dopplergrams that might be even filtered) at those points as $\Psi(\mathbf{r}_1, t)$ and $\Psi(\mathbf{r}_2, t)$. The key mathematical quantity to be evaluated is then the cross-covariance evaluated as

\begin{equation}
    C(\mathbf{r}_1, \mathbf{r}_2, t) = \frac{h_t}{T - |t|} \sum_{t'}\Psi(\mathbf{r}_1, t')\Psi(\mathbf{r}_2, t'+t)
    \label{eq:xcor}
\end{equation}
in its discrete form \citep{gizon-birch2002}. There $h_t$ is the temporal sampling and $T$ is the duration of observation.

\subsection{Travel times}
In realistic applications, the cross-covariance function measured for two points contains a large amount of realization noise due to the stochastic nature of solar oscillations. Hence, it is almost impossible to measure the travel time of the waves traveling between two individual pixels in the helioseismic observables. Usually, some averaging is required. Most often, such averaging is applied both in time and space. In time, cross-covariances are usually averaged over a time period $T$ larger than about 6~hours. To further improve the signal-to-noise ratio, it was first suggested by \citet{duvall1993} to average cross-covariance $C(\mathbf{r}_1, \mathbf{r}_2, t)$ over points $\mathbf{r}_2$ that belong to the annulus or quadrants centered at $\mathbf{r}_1$ with a distance $\Delta=|\mathbf{r}_2-\mathbf{r}_1|$. For the study of the waves travelling from the central point towards the surrounding annulus and in the opposite direction, averaging over a full annulus is suitable. For the investigation of the waves traveling in the direction parallel to the equator (east--west), averaging over 90-degree quadrants of the annulus in the corresponding direction is more suitable. In the methodology used in this study, quadrants were replaced by sections of the annulus multiplied by the cosine (east--west direction) or sine (south--north direction) of the horizontal polar angle. These geometries constitute a continuous transition from quadrant geometries.

For a given set of $\mathbf{r}_1$ and $\Delta$, the cross-covariance function oscillates around a characteristic travel time in both the positive and negative part of the time axis. It is a goal of the time--distance methods to search for the travel time that maximises the cross-covariance (\ref{eq:xcor}). One of the approaches to measure the travel time is to fit a Gaussian wavelet to the cross-covariance function \citep{1997ASSL..225..241K}, 

\begin{equation}
    C_+(t; \mathbf{r}_1,\Delta) = A\exp [ - {\gamma ^2}{(t - {t_{\rm{g}}})^2}]\cos [{\omega _0}(t - {\tau_+})],
    \label{eq:gabor_fitting}
\end{equation}
where $A$ is the amplitude, $t_{\rm g}$ represents the group travel time and $\tau_+$ the positive phase travel time. Parameters $A$ and $\gamma$ describe the amplitude and decay rate of the wavelet envelope. An analogous equation may be written for a negative travel time. The time--distance analysis relies on measurement and interpretation of the phase travel times $\tau_+$ and $\tau_-$.

The fitting of the Gaussian wavelet is demanding and often fails due to the presence of noise. Therefore, alternative definitions of travel times were developed that are more robust in the presence of noise. They stem from the fact that the cross-covariance of the waves may be computed from the solar model by solving the equations of the wave propagation. These cross-covariances represent some sort of a reference for a spherically symmetrical 1-D quiet Sun. We may safely assume that the changes to the cross-covariances caused by the perturbations of the model will introduce a modification of the background-model cross-covariances. Then, instead of fitting a 5-parameter Gaussian wavelet, we may use this reference cross-covariance as a template and perform only 1-parameter fitting in time. Such an approach is used also in geoseismology \citep{zhao1998} and for the need of the time--distance helioseismology was devised by \citet{gizon-birch2002}. Following the paper we refer to these travel times as to GB02 travel times henceforth. Note that in this approach not the full travel time $\tau$ is measured, but the deviation $\delta\tau$ of the travel time from the reference template.  

Later, \citet{gizon-birch2004} simplified this definition even further for the case of small perturbations. Such a robust definition allows the measurement of travel times averaged over a short time (e.g., around $T=2$ hours) and with a large spatial resolution. This linearised approach is termed GB04 henceforth. 

We note that both GB02 and GB04 definitions of the travel times result in a travel time that is a mixture of both the group and phase travel times. It is not straightforward to decouple the two. A proper choice of the reference cross-covariance $C^0$ is essential to minimize the contribution of the group travel time to the results. In some cases, a reference cross-covariance coming from the forward modeling does not have to be accurate enough. In those cases, a construction of the reference cross-covariance as a spatial average of different realizations over the field of view in the quiet-Sun regions may be a better choice.

\subsection{Travel-time sensitivity kernels}
Let us consider that the solar model is fully described by a multidimensional vector of quantities $Q^\alpha(\mathbf{x})$, where the index $\alpha$ lists the considered physical quantities. These quantities may, for instance, be vector flows, sound speed, density, adiabatic exponent, among others. In our study, we primarily focused on the flows $\mathbf{v}$. The quantities $Q^\alpha$ are functions of 3D spatial coordinates $\mathbf{x}$ and, in principle, also of time. However, we will only consider stationary models in terms of the averaging over the observation interval $T$.

Similarly to the discussion of the travel times, we may split the values of the quantities into the unperturbed background $Q_0^\alpha(\mathbf{x})$, which in most cases is represented by the background solar model, and the perturbation $q^\alpha(\mathbf{x})$.

To compute the linear adiabatic oscillations of the Sun, methodologies have been available for a very long time \citep{lynden-bell1967}. Such a computation is slow and expensive even when considering the computational power of today's computers. This approach is not viable when dealing with realistic problems. If we focus only on linear perturbations of the model, we may relate travel-time deviations $\delta\tau$ with perturbations of the solar model $q^\alpha$ by an equation.

\begin{equation}
    \delta\tau^a(\mathbf{r}) =  \sum\limits_\alpha \int_{\odot}{\rm d^2}{\mathbf{r}'} {\rm d}{z} \mathbf{K}_{\alpha}^a(\mathbf{r}' - \mathbf{r}, z) \cdot q^\alpha(\mathbf{r}', z) + n^a(\mathbf{r}) \ .
    \label{eq:traveltimesdef}
\end{equation}

The index $a$ represents the travel-time geometry and combines the choice of the $k-\omega$ filter, point-to-annulus or point-to-quadrant averaging, and the radius of the annulus $\Delta$. Measured travel times are subject to random realization noise $n^a$. The quantity $\mathbf{K}_\alpha^a$ is termed a \emph{sensitivity kernel} and represents a function that translates changes in the solar model into travel-time deviations. The assumption of small perturbations allows us to split the total travel-time deviation into individual contributions by individual perturbers $\alpha$. 

Sensitivity kernel functions are calculated numerically using the background solar model. Various approaches have been developed by different authors. The simplest approximation ignores the finite-wavelength effects and assumes the propagation of waves along their optimal rays \citep[see][and references therein]{1997ASSL..225..241K}. Finite-wavelength effects are considered by approximating wave propagation in terms of scattering. Single-source sensitivity kernels were proposed and computed, for example, by \cite{birch2000}. Sensitivity kernels considering randomly distributed sources were proposed and developed by \cite{gizon-birch2002}. Later, a linearised approach consistent with the definition of linearised travel times was proposed by \cite{gizon-birch2004}. Gizon \& Birch emphasized the importance of incorporating data processing steps—not only the $k-\omega$ filtering but also an accurate estimate of the telescope's point spread function—into the calculation of the kernels, ensuring they align with the measured travel times. For details of the kernel calculation, we refer to \cite{burston2015}.

It needs to be mentioned that there are other, more sophisticated methods allowing to compute sensitivity kernels, namely the adjoint method \citep{2011ApJ...738..100H}. The method relies on the computation of the wave field that is driven by two sources. One, by the excitation spectrum resulting in the forward wavefield, and, two, the difference between the predicted wavefield and observations (the adjoint wavefield). The method thus minimises the misfit between the modelled wavefield and the observed ones, whereas the sensitivity kernels are produces naturally within the process. 

\subsection{Motivation to this work}
In real applications, the problem is usually posed in the opposite direction: from the measured travel-time maps we want to learn about the properties of the solar interior, therefore we usually want to invert (\ref{eq:traveltimesdef}) to infer $q^\alpha$. Not mentioning issues with the chosen inversion schemes used by the community, the procedure critically depends on the accuracy of the sensitivity kernels. It is the kernels that represent the background model of the Sun, around which the problem is linearised and solved. Should the kernels suffer from issues, the recovered tomographic maps of the solar interior must be affected by those issues. 

Recently, one of us  indicated the inconsistencies in the special type of inversions for flows, where the technique of ensemble averaging was applied \citep{svanda2015}. This technique uses an assumption that if many realisation of the same phenomenon are averaged (it was the supergranular cells in that case), such an averaging decreases the noise term in (\ref{eq:traveltimesdef}). Then in the inversion procedure, term containing the sensitivity kernels is the most important and the results of such statistical inversion are constrained only by those kernels. The study showed that various choice of sensitivity kernels in the inversion lead to very different models of the average supergranule. It was not possible to tell from the results and travel times alone, which of the models is ``better'', that is closer to reality. 

Later, similar issue was confirmed by \cite{degrave2015}, who showed that the observed travel-time maps are inconsistent with the forward modelled travel times using various supergranular models. This study in fact directly tested Eq.~(\ref{eq:traveltimesdef}) and pointed out possible issues with the kernels. 

On the other hand, \cite{2006ApJ...646..553D} directly measured sensitivity kernels for flows for the surface gravity modes utilising small magnetic points as scatterers and obtained a reasonable match with the models. 

A direct validation of the sensitivity kernels with observations is not possible. It is possible to test the kernels with respect to the Sun-like MHD simulations, however, most of the available simulations are not exactly Sun-like in terms of the properties of the surface flows and spectrum of solar oscillations. It motivated us to test at least some of the properties of the kernels with observations.

\section{Testing procedure}
The aim of this study is to test the validity of velocity (flow) perturbation sensitivity kernel integrals with a model-independent method. As a tool, we inject an artificial constant longitudinal flow (velocity perturbation) $\mathbf{v}_0$ into data from observations. This is achieved by tracking the selected region on the solar disk with a (known) constant velocity different from the solar rotation. This effectively means that to all perturbations $q^\alpha$ mentioned in (\ref{eq:traveltimesdef}), an additional perturbation $\mathbf{v}_0$ representing the constant injected velocity is added, transforming the equation into

\begin{equation}
    \delta\tau^a(\mathbf{r}) =  \sum\limits_\alpha \int_{\odot}{\rm d^2}{\mathbf{r}'}{\rm d}{z}\mathbf{K}_\alpha^a(\mathbf{r}' - \mathbf{r}, z) \cdot [ q^\alpha(\mathbf{r}', z) - \mathbf{v}_0] + n^a(\mathbf{r})\ .
    \label{eq: TT perturbation with added velocity (base eq)}
\end{equation}

The reason there is a minus sign in the term $[ q^\alpha(\mathbf{r}', z) - \mathbf{v}_0]$ is that we inject the artificial flow by moving the coordinate frame. However, moving the frame in one direction corresponds to an apparent flow of the same velocity but in the opposite direction, similarly to watching a landscape passing by through the window of a moving train. The term with $\mathbf{v}_0$ in the integrand is bound with a corresponding vector sensitivity kernel where $\mathbf{K}_v^a$. Since $\mathbf{v}_0$ is constant and independent of position, it is not affected by the volume integral and can be taken out of it:

\begin{align}
    \delta\tau^a(\mathbf{r}) = & \sum\limits_\alpha \int_{\odot}{\rm d^2}{\mathbf{r}'}{\rm d}{z}\mathbf{K}_\alpha^a(\mathbf{r}' - \mathbf{r}, z) \cdot q^\alpha(\mathbf{r}', z) \notag \\
    & - \mathbf{v}_0 \cdot \int_{\odot}{\rm d^2}{\mathbf{r}'}{\rm d}{z}\mathbf{K}_v^a(\mathbf{r}', z) + n^a(\mathbf{r})\ .
    \label{eq: TT perturbation with added velocity - v out of integral}
\end{align}

Strictly speaking, the additional velocity $\mathbf{v}_0$ should not be mistaken for the change of the rotational rate of the reference frame, because it is constant with depth. A constant angular velocity corresponding to a different rate of solar rotation would manifest with a decreasing tangential velocity with depth. At the surface, the two approaches are equal. However, most of the time--distance kernels we have at our disposal are sensitive only in a few Mm depth below the surface, hence the inaccuracies from exchanging these two points of view on interpretation of $\mathbf{v}_0$ in (\ref{eq: TT perturbation with added velocity - v out of integral}) are negligible. What is not negligible at all is the contribution of the random-noise realisation $n^a$. A priori it also is not known. One of the ways to suppress this term is by means of the spatial averaging. At the position of neighbouring pixels, the random realisation is correlated \citep[e.g.][]{gizon-birch2004}, the correlation length-scale on the other hand is only a few pixels. Hence by averaging over many pixels (100$\times$100 in our case) the contribution of the random noise is significantly lowered. 

Equation (\ref{eq: TT perturbation with added velocity - v out of integral}) then becomes 

\begin{align}
    \left< \delta\tau^a(\mathbf{r}) \right> = & \left< \sum\limits_\alpha \int_{\odot}{\rm d^2}{\mathbf{r}'}{\rm d}{z}\mathbf{K}_\alpha^a(\mathbf{r}' - \mathbf{r}, z) \cdot q^\alpha(\mathbf{r}', z) \right> \notag \\
    & - \mathbf{v}_0 \cdot \int_{\odot}{\rm d^2}{\mathbf{r}'}{\rm d}{z}\mathbf{K}_v^a(\mathbf{r}', z)  \ ,
    \label{eq: tt perturbation averaged}
\end{align}
where angle brackets indicate the spatial averaging. The first term on the right side of Eq. \ref{eq: tt perturbation averaged} can be viewed as a certain average background travel time perturbation which is unaffected by the implanted velocity. The second term is independent of position, therefore the averaging will not affect it and it remains unchanged. It can be seen that the average travel time perturbation in a selected central area of the tracked region is directly proportional to the volume integral of the velocity perturbation kernel with the injected velocity acting as a constant of proportionality. The injected velocity has only a longitudinal (east--west, usually termed as $v_x$) component. Naturally, east--west quadrant-like difference travel times are most sensitive to this kind of perturbation \citep{burston2015}. Then, finally 

\begin{equation}
    \left< \delta\tau^a(\mathbf{r}) \right> = \left< \delta\tau_{\rm{back}}^a(\mathbf{r}) \right> -v_0\int_{\odot} {\rm d^2}{\mathbf{r}'}{\rm d}{z}K_{v_x}^a(\mathbf{r}', z)
    \label{eq: tt perturbation averaged final}
\end{equation}

By applying a set of implanted velocities $v_0$ we may independently measure the spatial integral of the sensitivity kernel $K_{v_x}^a$ using a simple linear fit stemming from (\ref{eq: tt perturbation averaged}).

\section{Data and Methods}

\subsection{Data}

The travel times were measured from a series of Dopplergram datacubes. Dopplergrams were observed by the HMI/SDO instrument \citep{2012SoPh..275..207S,2012SoPh..275..229S} and stored within the Joint Science Operations Center (JSOC\footnote{\url{https://jsoc.stanford.edu}}). HMI scans the Fe$^{\rm I}$ 617.3~nm photospheric absorption line at six positions and measures the intensity across different polarization states. The Dopplergrams are computed using a fast ``MDI-like algorithm'' \citep{2012SoPh..278..217C} that employs the use of the discrete approximations of the Fourier coefficients describing the spectral line and correction by factors coming from the pre-computed look-up tables. The final data products are stored within the series {\tt hmi.V\_45s} at JSOC, from which are read by our pipelines using a module in {\sc SunPy} package. 

\subsection{Tracking pipeline}

The tracking pipeline is responsible for preparing datacubes that are then used for measuring the travel times. We wrote a versatile pipeline from scratch using routines from {\sc SunPy} \citep{2020ApJ...890...68S} package so that it exactly matches our needs. In addition to the traditional steps such as selecting the local area, applying the azimuthal equidistant (Postel) projection, and creating the datacube in FITS format with a correct header, this pipeline offers an additional feature specific to this work: implanting artificial velocities by apparently moving the selected area at a predefined speed. This is a crucial feature used in determining the model-independent velocity kernel integrals as explained in previous sections. Since this is not part of standard pipelines for datacube preparation, our own solution was necessary.

The pipeline is highly versatile and prepared to be executed both on a single computer and in a cluster environment. Individual consecutive steps were written as separate modules, so the user has the ability to check the outcome of each step before proceeding further. At a high level, the tracking pipeline performs the following:

\begin{enumerate}
    \item Prepare the folder structure together with all necessary configuration files (both for the datacube creation step and for the subsequent travel-time pipeline) based on the input configuration (JSOC DRMS queries, data PATHs, projection origins, implanted velocities).
    \item (OPTIONAL) Download data (Dopplergrams) from JSOC. This step is unnecessary when the required data are already downloaded.
    \item Create a datacube for each combination of DRMS query and velocity and save it in the form of a FITS file into its designated folder prepared in step 1.
\end{enumerate}

The configuration of the pipeline is given in a form of a text file. Among others, it contains the DRMS request string specifying the time range to be downloaded and processed, the positions of the centre of the tracked region in Carrington coordinates, and lower and upper limits for implanted velocities, and the velocity sample count. The set of implanted velocities is then randomly sampled from within the given velocity bounds. Subsequenty, a loop is executed over the realisation of the implanted velocities. Within this loop, the full-disc Dopplergrams are downloaded (they are downloaded only once for the given DRMS query), tracked and mapped following the requests, and stacked to a datacube. 

Before a final saving into a FITS file with headers compatible with the following processing, invalid values (NaNs or blank pixels) are removed from the datacube frames and replaced by a median of non-empty values. Then, from each frame, a quadratic smooth surface is removed to subtract large-scale trends in Dopplergrams. Such a large-scale trend smoothly varying across the field of view is naturally produces e.g. by the solar rotation. 

\begin{figure}
    \includegraphics[width=0.5\textwidth]{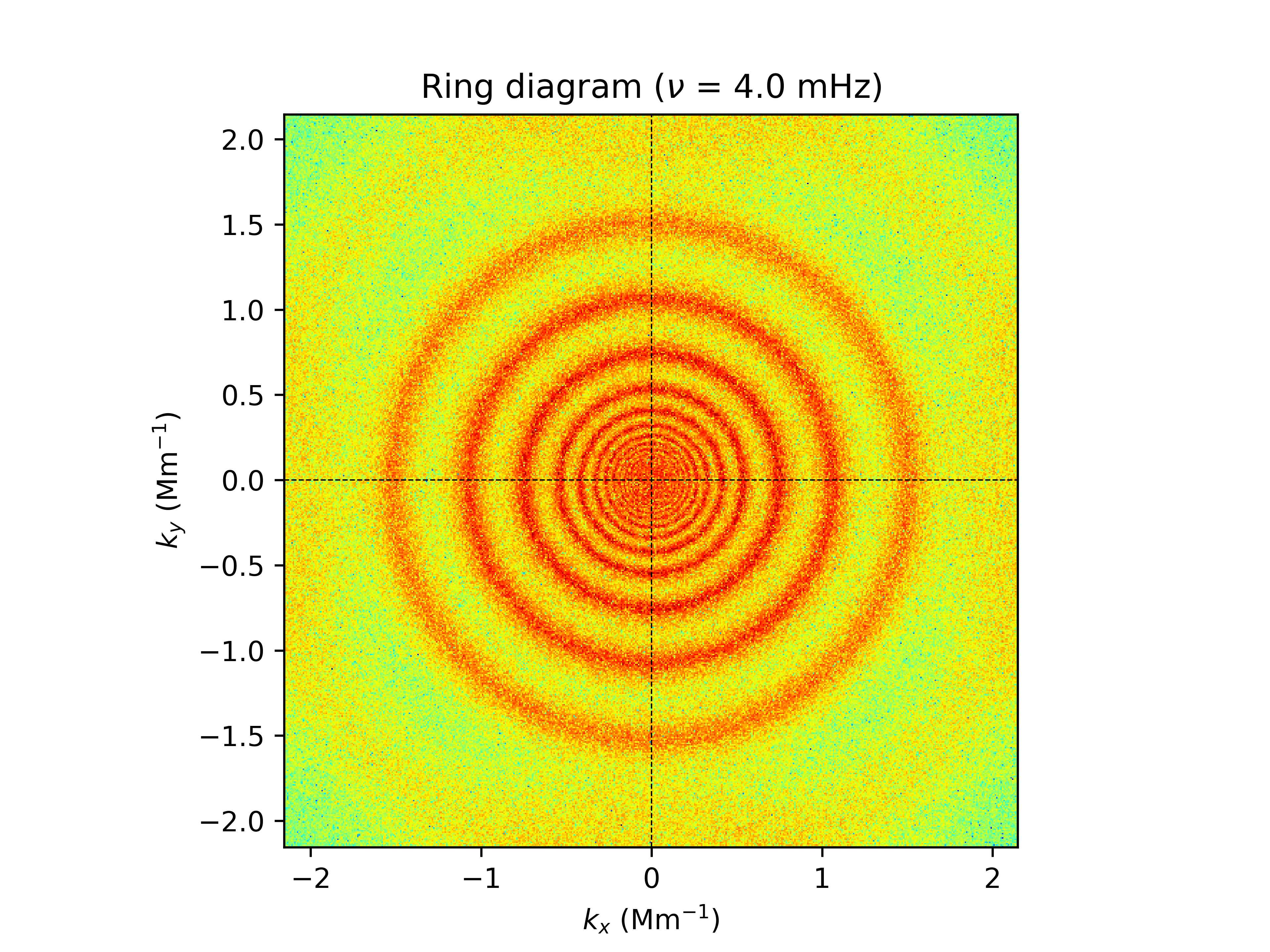}\\
    \includegraphics[width=0.5\textwidth]{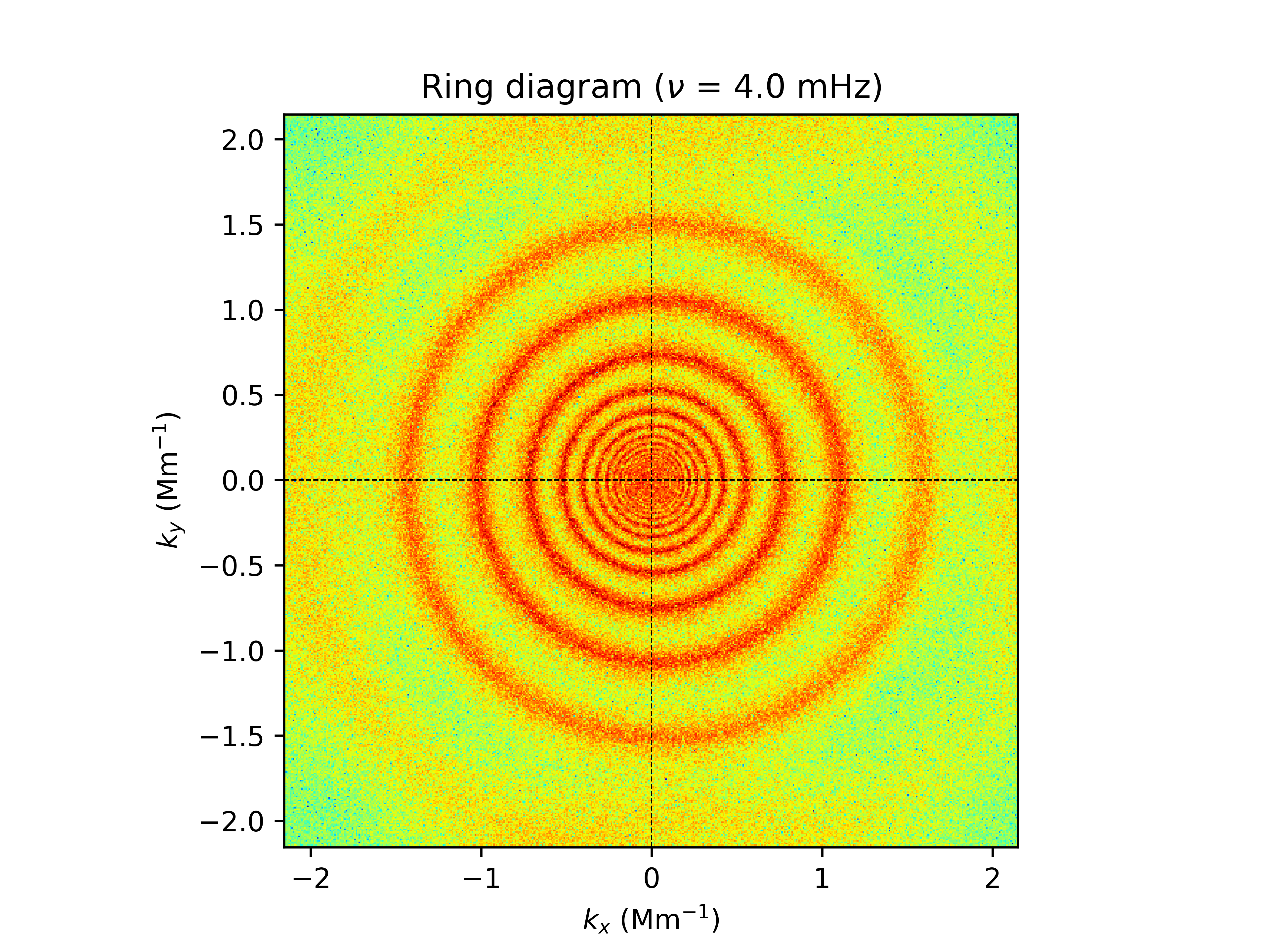}
    \caption{Cuts through the power spectra at a frequency of 4~mHz in the datacube that was tracked with a solar rotation (upper panel) and the same datacube tracked 500~m/s faster than solar rotation. The distortion of the rings due to the artificial velocity is clearly seen. }
    \label{fig:rings}
\end{figure}

\begin{figure}
\end{figure}

One potential question concerns whether the injected longitudinal velocity truly results in an artificial flow within the data. To address this, ring diagrams were constructed from the datacube power spectrum that show that the rings are shifted in the datacubes with an implanted velocity with respect to the situation in the datacube that was tracked with the solar rotation (see Fig.~\ref{fig:rings}). It qualitatively demonstrates the presence of a horizontal flow in the east--west direction, as evidenced by the horizontal distortion of the rings.

A direct comparison of the horizontal velocity with the injected one is also possible. In the Dopplergrams, the indication for the large-scale convective cells of supergranulation \citep{rincon2018} are clearly visible. The supergranular structures therefore can be used to determine the horizontal flow field on the solar surface \citep[e.g.][and the follow-up papers]{svanda2006}. Such goal may be achieved by applying the method of local correlation tracking \citep[LCT;][]{november1986} to Dopplergram datacubes. 

The local correlation tracking algorithm searches for the optimal displacement that minimises the differences of the floating spatial windows capturing slowly evolving features in the time sequence. Knowing the sampling of the cross-correlated frames, the detected displacement may be converted to the horizontal velocity vector. In the test performed here, the horizontal velocity vector is dominated by the longitudinal component, which is the only one we compare with the implanted velocity. 

\begin{figure}
    \centering
    \includegraphics[width=0.5\textwidth]{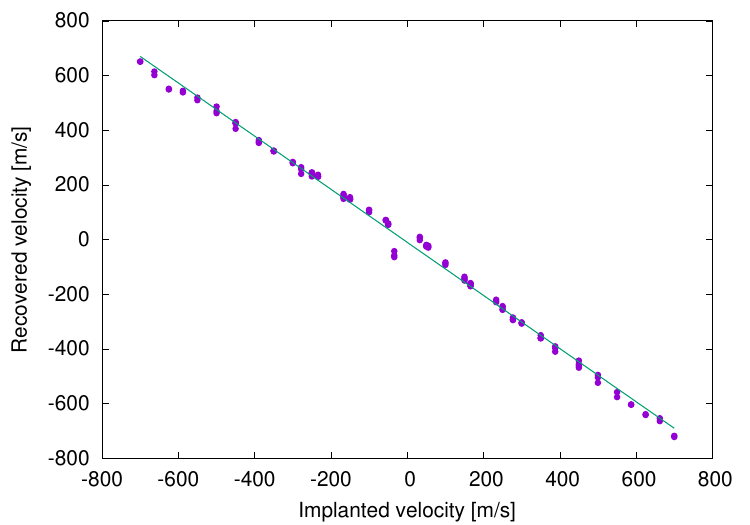}
    \caption{Comparison of the implanted velocity and the longitudinal component of the horizontal velocity vector determined from the manipulated datacubes using a LCT algorithm. The fit has a slope of $-1$, which is a consequence of the fact that the implanted velocity was injected by artificially changing the frame of reference, which naturally introduced an apparent flow in the opposite direction. }
    \label{fig: LCT}
\end{figure}

The results of the testing of our pipeline are shown in Fig.~\ref{fig: LCT}. The measured surface horizontal velocities obviously very well correspond to the implanted velocity, thereby validating the performance of the tracking pipeline. 

\subsection{Travel-time pipeline}
In general, measurements of helioseismic travel times consist of three consecutive steps. The travel-time pipeline operated on the data-processing computer cluster at the Astronomical Institute of the Czech Academy of Sciences in Ondřejov is no exception. First, the input datacube—the output of the tracking pipeline—is detrended in time, and a mean frame (average of the frames over time) is subtracted. This step removes existing large-scale trends in the observations. In Dopplergrams, such large-scale trends may be produces e.g. by a residual signal from the differential rotation and also by long-living velocity features, like most of the signal from solar supergranules. The detrended datacube is apodized in both space and time and transformed to Fourier space for further processing. 

Fourier-transformed datacube is filtered to keep only certain types of waves. We use two kinds of filtering traditional in time--distance helioseismology, that is the ridge filters (isolating only the surface gravity $f$ modes or acoustic $p$ modes to the fourth order) and the phase-speed time--distance filters \citep[indicated as td1 to td11, following][]{1997SoPh..170...63D}. 

These filtered datacubes are inputs for the travel-time measurements procedure. Following the usual recipes, cross-covariance maps are computed numerically (see Equation (\ref{eq:xcor})) for spatially smoothed datacubes utilising the generalised point-to-annulus geometry. Then, the travel times that minimize the cross-covariance of the signals at the given point are measured using a use-selected definition. In our case, we used both GB02 and GB04 definitions. 

The above-described workflow was implemented in a pipeline coded in the MATLAB language. Each of the three steps is represented in a separate module and the pipeline is executed in the cluster environment. 

\subsection{Sensitivity kernels}
The sensitivity kernels that were tested were computed by {\sc Kc3} code by Aaron Birch \citep{2007AN....328..228B}. The code uses a mode summation approach in the Cartesian geometry and a Born approximation with randomly distributed sources. It uses eigenfunctions and eigenfrequencies of the oscillations from the solar model S \citep{1996Sci...272.1286C}. {\sc Kc3} is a reference kernel code that produced sensitivity kernels for flow or sound-speed perturbation that were used in many studies before. 

\section{Results}
To achieve the goals of this study, we analysed in total 716 separate datacubes. An overview is given in Table~\ref{tab:datacubes}. The processed data volume consisted of 705~GB in the Dopplergram datacubes and 8.5~TB in the resulting travel-time maps and necessary intermediate data products. The processing took about 2 months using a Radegast cluster at the Solar Department of the Astronomical Institute of the Czech Academy of Sciences in Ond\v{r}ejov, which has in total 102 CPU cores. 

For each ridge filter (f and p$_1$ to p$_4$ modes) we evaluated travel times for 16 different distances $\Delta$, starting at 5~px and ending at 20~px, with a pixel size of 1.46~Mm. The reason for the degraded MDI-like resolution is that the sensitivity kernels we have at our disposal and that we are regularly using for helioseismic inversions within our group \citep[e.g.][]{2013ApJ...771...32S,2013ApJ...775....7S,2014ApJ...790..135S,2019A&A...622A.163K,2021A&A...646A.184K} have exactly this pixel size. As for the phase-speed filters td1 to td11, we use 5 distances $\Delta$ for each, following the usual definition of these filters by \cite{couvidat2006optimal} \citep[also Table~1 in][]{2013ApJ...775....7S}. For each mode (filter) and distance we calculated travel-time maps for all three point-to-annulus and point-to-quadrant geometries. In the following, we only used the east--west averaging geometry, because it is the most sensitive to the zonal flows \citep{burston2015} we artificially introduce. Hence, we verified kernel integrals for 135 different sensitivity kernels in total. 

\begin{table}[!ht]
\caption{A summary table of the analysed datacubes. Each of the cubes covered 6 hours of observation and at the cadence of 45~s per frame consisted of 482 frames. The Carrington longitude $l$ and latitude ($b$) of the central pixels is also given. Note that except for the last two, all datacubes tracked the regions around the solar equator. The last column indicates the number of various tracking rates for the datacube at the given date. }
\begin{tabular}{lrrrrrr}
\hline
\rule{0pt}{12pt}Date & $l$ [deg] & $b$ [deg] & \# datacubes \\
\hline
\rule{0pt}{12pt}03/12/2017 & 280 & 0  & 54 \\
05/12/2017 & 200 & 0  & 74 \\
08/13/2017 & 40 & 0  & 75 \\
10/08/2017 & 30 & 0  & 75 \\
11/03/2017 & 40 & 0  & 75 \\
12/01/2017 & 40 & 0  & 75 \\
03/26/2018 & 310 & 0  & 75 \\
01/20/2019 & 315 & 0  & 75 \\
\hline
01/20/2019 & 315 & $-$20  & 63 \\
01/20/2019 & 315 & 20  & 75 \\
\hline
\end{tabular}
\label{tab:datacubes}
\end{table}

\begin{figure}[!ht]
    \includegraphics[width=0.5\textwidth]{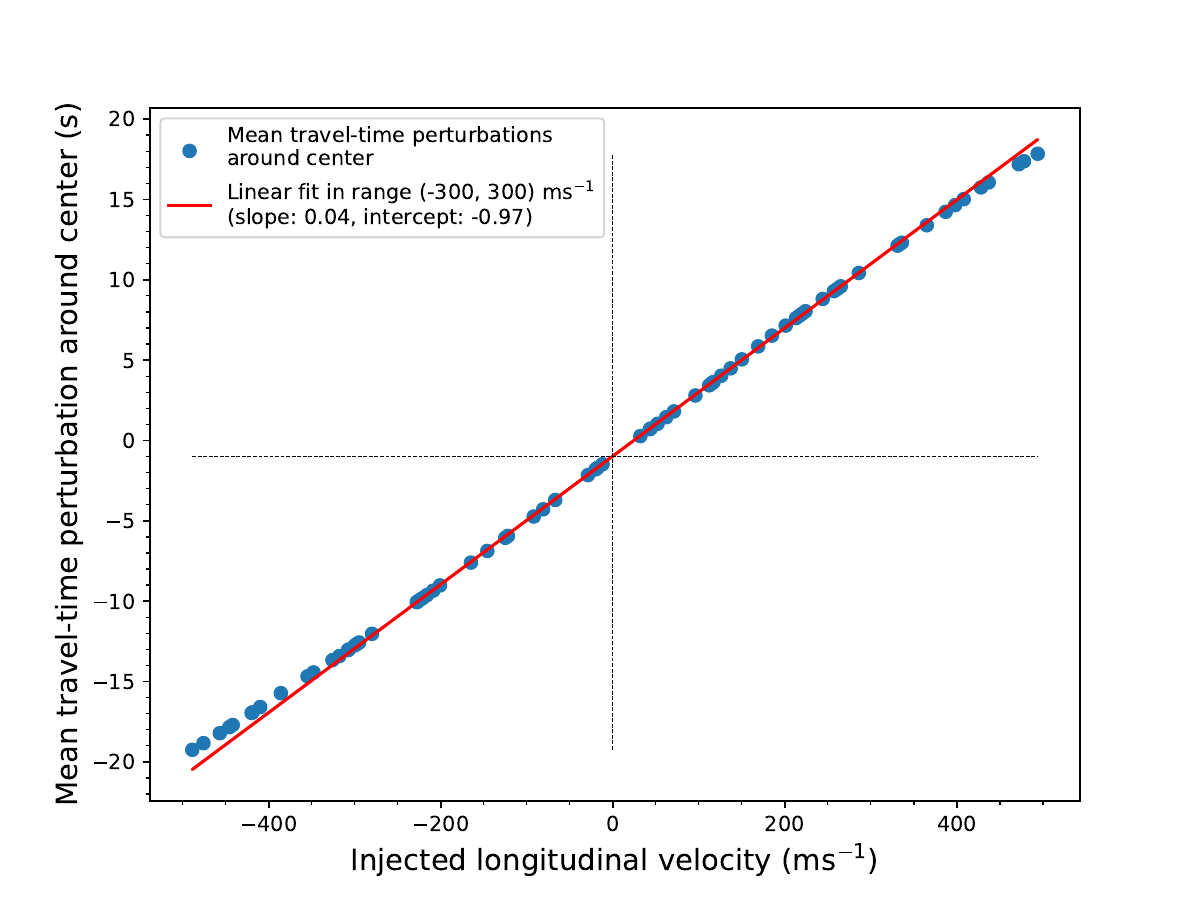}\\
    \includegraphics[width=0.5\textwidth]{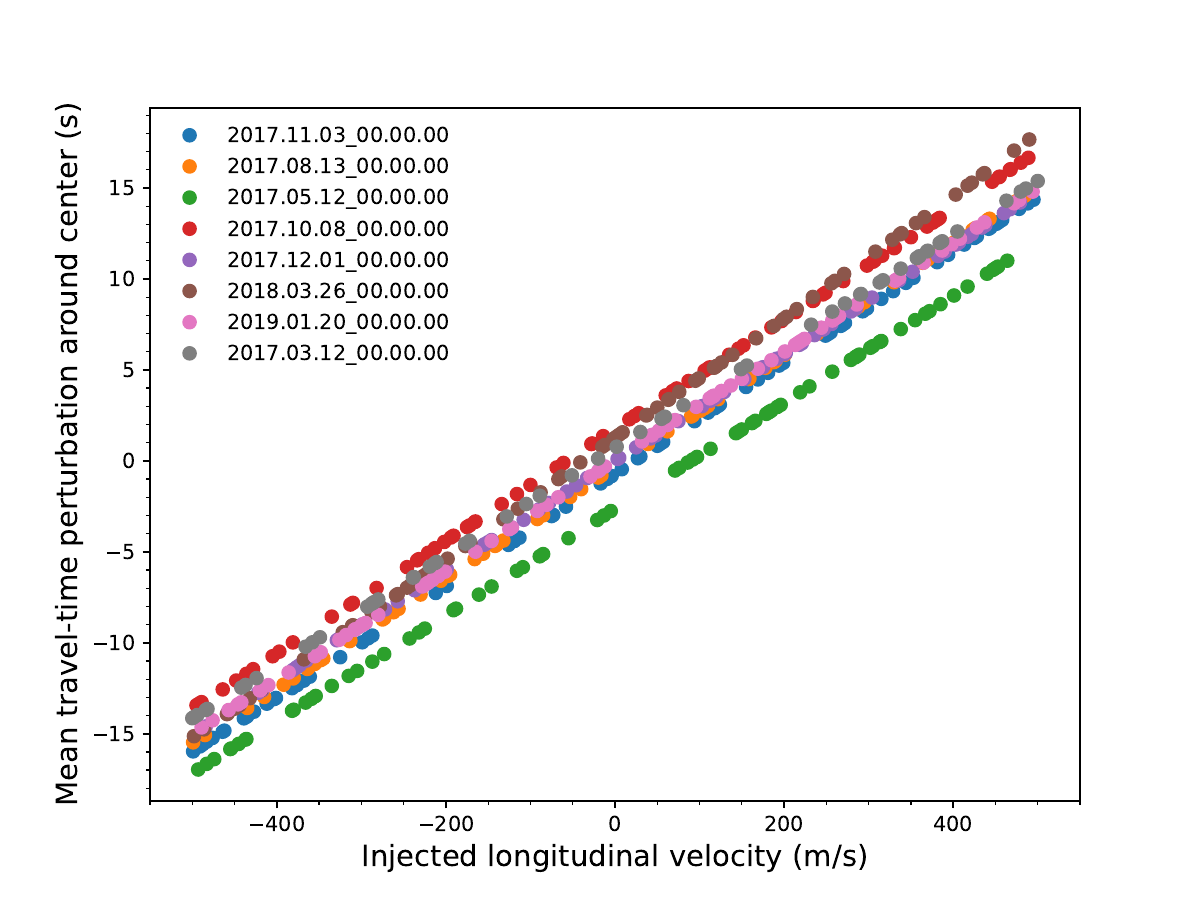}
    \caption{Upper: Measured average travel time around the disc centre for mode td4, distance of 9.9~px as a function of the implanted velocity. Linear fit and its properties are also indicated. The slope of the linear fit corresponds to the total integral of the appropriate sensitivity kernel. Bottom: Similar plot for the p$_3$ mode and $\Delta=15$~px when using several dates. Different dates group around virtual lines that are shifted vertically, as indicated by time stamps. }
    \label{fig:implaneted_vs_meantt}
\end{figure}

A procedure is demonstrate e.g. in Fig.~\ref{fig:implaneted_vs_meantt}. There, for a chosen sensitivity kernel (index $a$), we plot the dependence given by Eq.~(\ref{eq: tt perturbation averaged final}), that is the dependence of the average travel-time deviation in the central region $\left< \delta\tau^a(\mathbf{r}) \right>$ of the datacube as a function of the implanted zonal velocity $v_0$. The dependence is monotonous (that is the case for all considered $a$s) and very close to linear, as expected. For larger values of $v_0$ above say 300~m/s, the average travel times start to deviate slightly from the linear fit, which is also overplotted. Such a non-linearity is expected for large velocities and is consistent with previous works \citep[see e.g.][]{2007ApJ...671.1051J}. Therefore, we evaluated the fit of the linear function only in the range of $(-300,+300)$~m/s. For higher modes (e.g. $p_4$) and larger phase speeds the non-linearity is slightly more pronounced. Following Eq.~(\ref{eq: tt perturbation averaged final}) the slope of the linear fit corresponds to the total spatial integral of the corresponding sensitivity kernel. 

For the different dates, a different realisation of the background travel-time term $\left< \delta\tau_{\rm{back}}^a(\mathbf{r}) \right>$ in (\ref{eq: tt perturbation averaged final}) occurs (see Fig.~\ref{fig:implaneted_vs_meantt} bottom panel). For instance, a uncorrected differential rotation will contribute to this term. This term changes the intercept of the linear dependency. We are interested only in the slopes. In order to learn about the characteristic slope value, we fit the linear dependence individually for each date and then obtain the representative slope value $s$ as an weighted average:
\begin{equation}
    s = \frac{\sum\limits_i w_i s_i}{\sum\limits_i w_i}\ ,\quad {\rm with }\quad  w_i = \frac{1}{\sigma_{s,i}^2}\ ,
\end{equation}
where $s_i$ are the individual slopes with their error-bars $\sigma_{s,i}$. 

\begin{figure}[!ht]
    \includegraphics[width=0.5\textwidth]{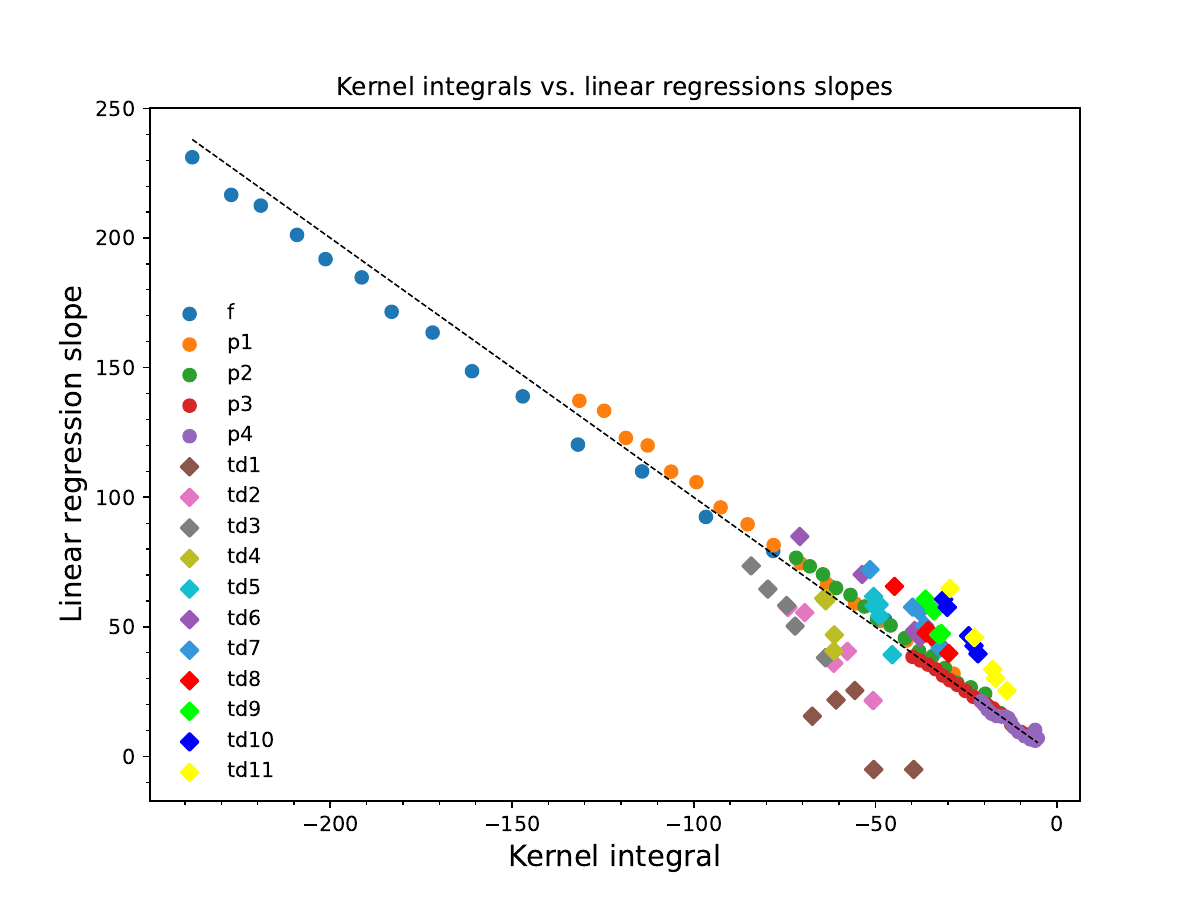}
    \caption{Volume integrals of velocity perturbation kernels obtained from linear regression slopes (model independent) vs. volume integrals of the velocity perturbation kernels calculated from forward modelling. The data were obtained from 578 different datacubes for various ridge and phase-speed filters. The dashed line indicates the expected line with a slope of $-1$. Values on both axes are in units of s/(km/s).}
    \label{fig:measured_vs_model}
\end{figure}

The comparison of the resulting slope that represents the model-independent value of the total spatial integral of the sensitivity kernel with the value from the forward modelling for all modes and all distances is plotted in Fig.~\ref{fig:measured_vs_model}. Colours represent different modes, the total integral values usually increase with the distance $\Delta$. An expected line with a slope of $-1$ is overplotted for reference. 

From the figure it can be seen that the measured and modelled travel-time sensitivity kernel integrals correspond to each other for most of the ridge filters. This is also demonstrated further in Table~\ref{tab:slopes}, where the slopes for the f and p$_1$ to p$_4$ modes are close to the expected value of $-1$. The fits show a small spread, small intercept values, large (anti)correlation, and are statistically significant. The slope deviates from $-1$ for the p$_4$ mode, which penetrates to deeper layers, where the assumption of implanting the surface velocity that is applicable to the whole sub-surface region does not have to hold any longer.

\begin{figure}[!ht]
    \includegraphics[width=0.5\textwidth]{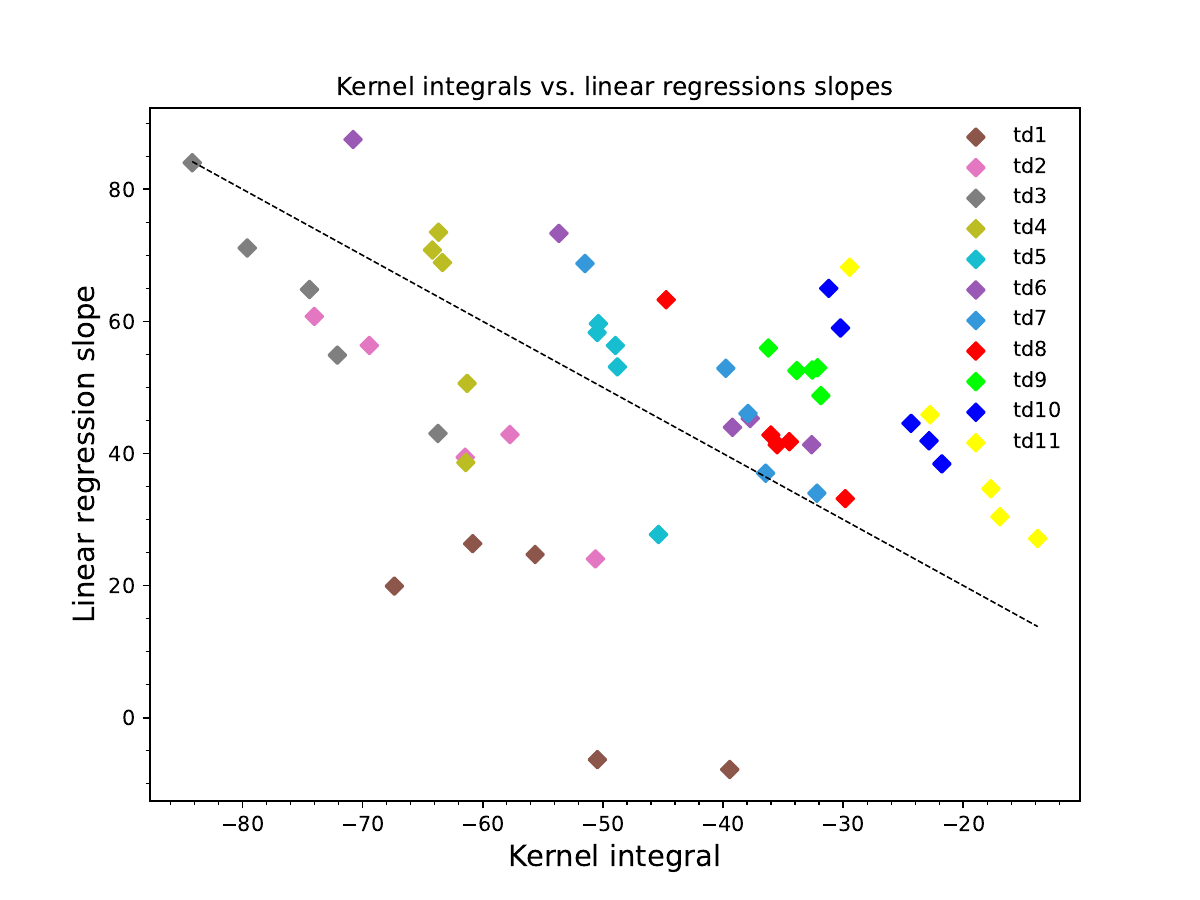}
    \caption{Same as Fig.~\ref{fig:measured_vs_model} only with details to phase-speed filtered kernels. Also, this plot is obtained for a particular date, not the combination of more. The dashed line indicates the expected line with a slope of $-1$. Values on both axes are in units of s/(km/s).}
    \label{fig:measured_vs_model_TD_only}
\end{figure}

The story is different for the phase-speed filters (see e.g. Fig.~\ref{fig:measured_vs_model_TD_only}). Even when the correlation coefficient is close to $-1$ and the fits are significant (all cases except for td1 and perhaps td4, when we take the level of statistical significance 0.01), the slopes differ from the expected value of $-1$ significantly. Also, the values of the intercepts are mostly large. The most notable differences can be seen for low phase speeds, for filters say td1 to td5. Large deviations from the expectations are registered also for the highest phase speeds (filters td9 to td11), where the slope is much larger than the expected value. Some of this disagreement is probably due to the reason already stated above when discussing the p$_4$ mode. On the other hand, p modes with low phase speeds only penetrate shallow near-surface layers and our testing approach should be valid for them. 

We note that the linear fits by mode are rather indicative. A complete set of kernel integrals from the forward modelling and their comparison to the model-independent determination is given in Table~\ref{tab:allkernels}. 

\begin{table*}[!ht]
    \caption{This table indicates the properties of the linear fits comparing the derived kernel integrals with the modelled values for individual modes. For each mode, a set of distances $\Delta$ was used. The modes that behave according to the expectations have slopes around value of $-1$, a small value of intercept and correlation coefficient $\rho$ around $-1$. Also, the $p$ value of the fit is significant ($0.01$ or less).}
    \centering
    \begin{tabular}{l|cccccc}
    \hline
    \rule{0pt}{12pt}Mode & Slope & $\sigma_{\rm Slope}$ & Intercept & $\sigma_{\rm Intercept}$ & $\rho$ & $p$-value \\
    \hline
\rule{0pt}{12pt}f & $-$0.93 & 0.02 & 4.02 & 2.57 & $-$1.00 & $<0.01$\\
p$_1$ & $-$1.04 & 0.01 & 1.39 & 0.85 & $-$1.00 & $<0.01$\\
p$_2$ & $-$1.06 & 0.01 & 1.31 & 0.70 & $-$1.00 & $<0.01$\\
p$_3$ & $-$0.97 & 0.01 & 0.67 & 0.33 & $-$1.00 & $<0.01$\\
p$_4$ & $-$0.89 & 0.07 & 1.56 & 0.96 & $-$0.96 & $<0.01$\\
td1 & $-$1.0 & 0.55 & $-$44.22 & 30.58 & $-$0.72 & 0.17\\
td2 & $-$1.53 & 0.26 & $-$53.40 & 16.66 & $-$0.96 & 0.01\\
td3 & $-$1.74 & 0.11 & $-$72.87 & 8.12 & $-$0.99 & $<0.01$\\
td4 & $-$6.67 & 1.37 & $-$364.85 & 86.14 & $-$0.94 & 0.02\\
td5 & $-$4.14 & 0.64 & $-$147.70 & 31.38 & $-$0.97 & 0.01\\
td6 & $-$1.16 & 0.09 & 4.39 & 4.41 & $-$0.99 & $<0.01$\\
td7 & $-$1.50 & 0.18 & $-$3.74 & 7.33 & $-$0.98 & $<0.01$\\
td8 & $-$1.76 & 0.13 & $-$14.05 & 4.65 & $-$0.99 & $<0.01$\\
td9 & $-$3.40 & 0.59 & $-$61.70 & 19.63 & $-$0.96 & 0.01\\
td10 & $-$2.12 & 0.09 & $-$5.96 & 2.29 & $-$1.00 & $<0.01$\\
td11 & $-$2.58 & 0.13 & $-$11.89 & 2.62 & $-$1.00 & $<0.01$\\
\hline
    \end{tabular}

    \label{tab:slopes}
\end{table*}

\section{Discussion}
Our results indicate an unresolved issue with the travel-time sensitivity kernels utilising phase-speed filtering. Some of the disagreement might be a consequence of the chosen testing procedure, where the implanted velocity is under control only at the surface. Deeper down, its proper interpretation becomes more difficult. The waves with large phase speeds and higher modes are more sensitive to the state of deeper, this property is prominent especially for the phase-speed-filtered waves \citep[see e.g. Fig.~2 in][]{2013ApJ...775....7S}. Therefore, it is difficult to properly interpret the disagreement between the modelled and observed travel-time sensitivity kernel integrals for those modes. On the other hand, the disagreement is also present for very shallow waves (short distances in ridge-filtered data and lowest phase-speed filtered data with modes td1 to td4), where high-degree modes prevail. The disagreement of amplitudes and line widths of the models for the high-degree modes with the observations is known from the literature \citep[see e.g.][]{2013ApJ...772...87K}, which affects the Green's functions and cross-covariance functions used in the kernel models, among others. 

It would seem at the first sight that the different conclusions drawn for ridge- and phase-speed-filtered kernels might indicate different approaches taken in processing the ridge- and phase-speed-filtered data. We want to stress out that this is not the case. In our testing procedure, identical kernel code and identical travel-time pipeline was used to process different filtering approaches. The corresponding $k-\omega$ filters are supplied to the code as external files having the same format. Both codes/pipelines are written in {\sc Matlab} programming language and share as many common sub-routines as necessary. We would like to stress out that our travel times and sensitivity kernels are mutually consistent in terms of the considered data processing steps. Filtering, spatial resolutions, etc. are taken into account consistently in both the computation of the sensitivity kernels and measuring the travel times. 

There is a hand full of choices in running the travel-time pipeline that might possibly influence the results of the presented comparison. 

\paragraph{Choice of the travel-time fitting method}
Our travel-time pipeline offers in total three definitions of the travel time to be utilised: Gabor-wavelet fitting, GB02, and GB04. We only used GB02 and GB04 in this study, because our travel-time sensitivity kernels are compatible with those definitions. We tested that both approaches provide us with essentially indistinguishable results for implanted velocities in the range of about $(-200,200)$~m/s, then non-linearities start to appear for GB04 definition for some travel-time geometries. This is no surprise as GB04 definition is only valid for small perturbations. For the GB04 definition, the points start to deviate from the linear fit sooner than as shown e.g. in Fig.~\ref{fig:implaneted_vs_meantt}, which was computed with the GB02 definition. For this reason, despite being computationally significantly more demanding, we preferred the GB02 definition for our final results, because it allowed a larger range of the implanted velocities to be examined. 

\paragraph{Averaging time}
Averaging of the cross-covariances over a larger time span leads to a lower contribution of the random noise, the signal-to-noise ratio scales approximately as $\sqrt{T}$ with $T$ being an averaging time. We computed several testing runs using $T=24$~hours and $T=6$~hours and compared the results. Averaging of the travel times in the spatial window around the centre of the tracked patch (we used $100\times 100$~px window) is much more effective in suppressing the random noise than choice of the larger averaging time. Hence the averaging time has no effect on the results of this study. 

\paragraph{Choice of the reference cross-covariance}
Both GB02 and GB04 definitions of the travel times rely heavily on the reference cross-covariance, which is used as a template when finding the offset $\delta\tau$ of the measured travel time. As pointed out already in the introduction, the reference cross-covariance should be close to the measured one. In the case of the small implanted velocities, a usual quiet-Sun reference cross-covariance may be used, e.g. that is coming from the forward modelling. With increasing $v_0$, the deviations from the template became significant and break the assumption of the perturbation, that is a small change. One has to realise that when tracking the surface at a rate significantly differing from the background solar rotation, even position of the emergent point changes, which again influences the measured cross-covariance when applying the non-modified pipeline. E.g. for the wave travelling about 20~minutes from the central point to the surrounding annulus (which is a characteristic time of the propagation in our choice of the waves) the emergent point shifts by about 600~km with a tracking speed of 500~m/s with respect to the background solar rotation. This is about 0.4~pixels, which may already play a role in the measurement procedure. 

We found that the use of the pre-computed cross-covariance from the solar model is insufficient to represent the changes introduced by the tracking. The use of the reference cross-covariance computed as an average over all representations in the field of view solves the issue. 

\paragraph{Heliographic latitude}
The results presented above were obtained with datacubes that had centres positioned on the solar equator. We ran the same procedure also for two datacubes that were located at heliographic latitudes of $+20$ and $-20$~degrees. The results were very similar, they only differed by the value of the intercept on the implanted velocity vs. averaged travel time plot (see Fig.~\ref{fig:implaneted_vs_meantt}). The shift of the intercepts is a consequence of the non-compensated differential rotation, as the basic rotation rate was the Carrington rotation rate. The comparisons between the measured sensitivity kernel integrals and the models were comparable to those presented in Table~\ref{tab:slopes}. 

\begin{figure}
    \centering
    \includegraphics[width=0.5\textwidth]{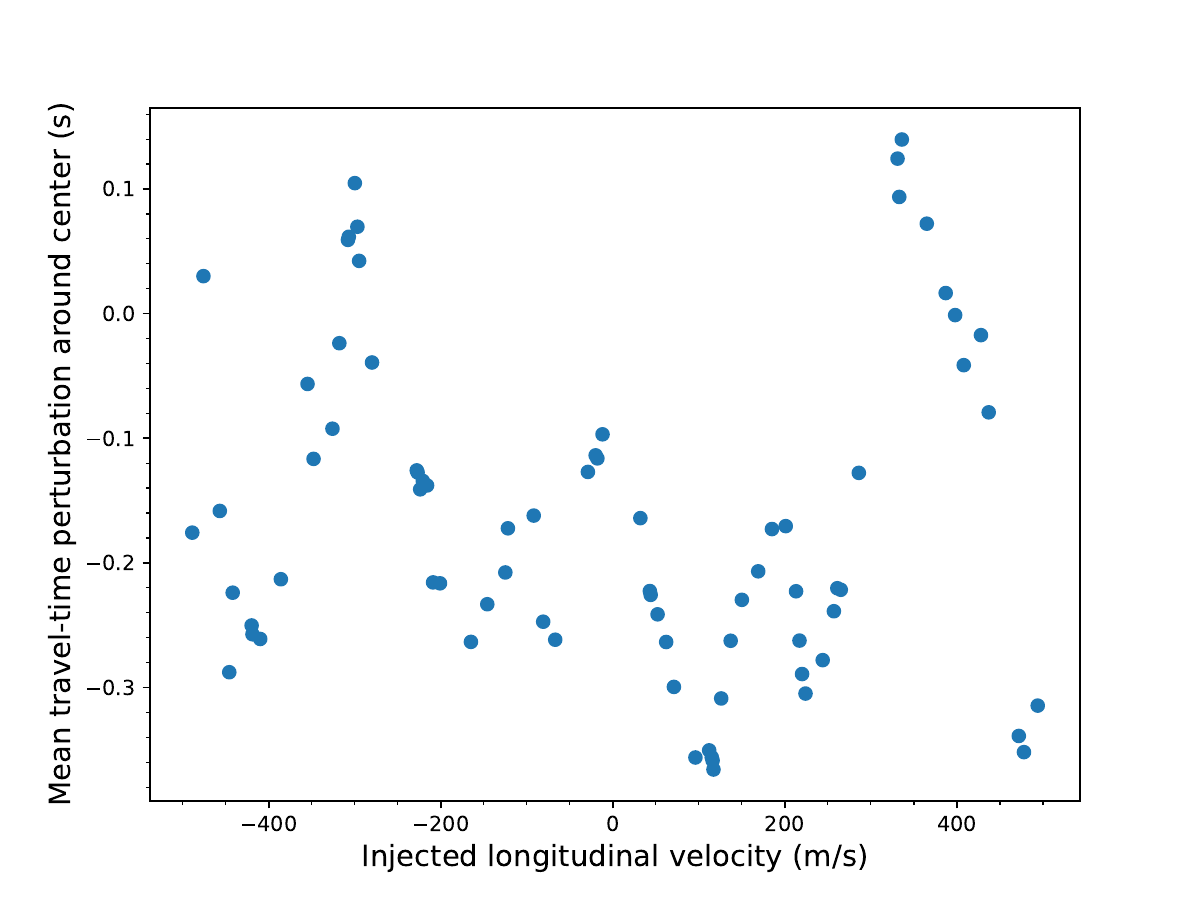}
    \caption{Figure similar to Fig.~\ref{fig:implaneted_vs_meantt}, only for the f mode, point-to-annulus geometry with $\Delta=11$~px. }
    \label{fig:oi}
\end{figure}

\paragraph{Other travel-time geometries}
In this study, we focused primarily to the east--west geometry from the point-to-quadrant choices, because it is naturally sensitive to the implanted zonal velocity \citep{burston2015}. Other geometries, such as the point-to-annulus (also known as out-minus-in) or south--north may be used. The volume integrals of those geometries are zero for the zonal ($v_x$) velocity. In agreement, the averaged travel times around the disc centre have a small value and vary quasi-randomly with implanted velocity (see Fig.~\ref{fig:oi}).


\section{Concluding remarks}
Our study testing the volume integrals of the time--distance sensitivity kernels for flows points out a possible issue in the sensitivity kernels utilising the phase-speed filtering scheme. We need to stress out that our study does not have to necessarily invalidate results and interpretations of the published helioseismic inversions in the past. Namely:
\begin{itemize}
    \item This test represents a form of a necessary condition sensitivity kernels should fulfil, but not a sufficient condition, as the procedure only tests the volume integrals of the kernels. 
    \item The disagreement of the total volume integral may lie in details that might be in fact marginal for the proper analysis. For instance, if there is a slight disagreement of the kernel value at pixels that are far from the central point and such disagreement has a large area, even when the difference is small, due to the 2-D integration it will affect the total integral by a large amount. 
    \item Similarly, a shifted mean horizontal value will behave the same. A small shift, yet over a large area will affect significantly the total integral of the sensitivity kernel.
    \item \cite{svanda2015} speculated that the reason in the inconsistency of the flow models derived from the inversion using different combinations of travel-time geometries was in the fact, that during the inversion, contributions of some kernels were cancelled by contribution of the others. Subtraction to zero critically depends on precise values and should the values have small errors, the results will be largely unconstrained. In the case when the combination leads to the summation of the kernels, such results should be more robust in presence of inconsistencies. Should this hypothesis turn out to be true, the inversion schemes then call for the method how to determine the optimal combination of the sensitivity kernels to be used in the scheme that would avoid large-scale subtractions. 
    \item Comparisons between different codes computing the sensitivity kernels were made in the past \citep[e.g.][]{2010AAS...21640009P,2016ApJ...824...49B} and they resulted in a generally good match between different techniques. The disagreement between various codes was to within a few per cents. Even the comparison of total integrals \citep{2010AAS...21640009P} of kernels using very different wave-field approximations (ray approximation vs. Born approximation) showed an acceptable agreement. It should be possible with numbers in Table~\ref{tab:allkernels} to verify the kernel integrals also using other codes, provided that data processing involved in the kernel computation is consistent with the travel-time measurements presented in this study. 
    \item We would also like to point out that except for the phase-speed filters td1--td4, the measured kernel integrals are somewhat proportional to the values obtained from the forward models (see Fig.~\ref{fig:measured_vs_model_TD_only}). This is merely an observation and no conclusion is drawn from it at the moment. 
\end{itemize}

\noindent A great deal of the helioseismic inferences was achieved using methods involving sensitivity kernels. Our test indicates that there might be an issue with some of them, therefore more work on verification and validation is needed. It should be mentioned, that there are also helioseismic methods that do not require the sensitivity kernels which use the full-waveform information about the solar oscillations \citep[see e.g.][]{2007ApJ...668.1189W, 2021ApJ...910..156H,2022ApJ...926..127M, 2024NatAs.tmp..119H}.

\begin{acknowledgements}
      This paper presents the results obtained within completing the MSc. studies of D.Ch. under the supervision of M.\v{S}. at the Faculty of Mathematics and Physics, Charles University. This work was partly done with a support of the institutional support ASU:67985815 of the Czech Academy of Sciences. M.\v{S}. further acknowledges the support by Czech-German common grant, funded by the Czech Science Foundation under the project 23-07633K and by the Deutsche Forschungsgemeinschaft under the project BE 5771/3-1 (eBer23-13412). We are grateful to Aaron Birch for making his kernel code free to use. 
\end{acknowledgements}

\bibliographystyle{aa}

\appendix
\onecolumn

\section{All investigated kernels}
\begin{longtable}{lr|rrr}
\caption{The following table contains considered numbers for each investigated sensitivity kernel. We state mode and distance $\Delta$, the volume integral of the kernel from the forward modelling, and the slope of Eq. (\ref{eq: tt perturbation averaged final}) that should correspond to the kernel integral. These two recent values are displayed in Figs.~\ref{fig:measured_vs_model} and \ref{fig:measured_vs_model_TD_only}. We recall that due to the implementation of the tracking, the two expressions for the volume integral of the kernel should have opposite signs. In the last column, the ratio of the two previous numbers is given. The ratios are colour-coded. When black, the difference between the numbers is less than 10\%, if \textcolor{blue}{blue}, the difference is less than 20\%. In the case of the \textcolor{orange}{orange} color, the difference is less than 50\% of the expected volume integral and if this threshold is exceeded, the ratio is colour-coded with a \textcolor{red}{red} colour.} \\
\hline
\rule{0pt}{12pt}Mode & $\Delta$ & Kernel integral & Slope of Eq. (\ref{eq: tt perturbation averaged final}) & Ratio \\
 & [px] & [s/(km/s)] & [s/(km/s)] &  \\
\hline
\rule{0pt}{12pt}
f & 5.0 & $-$47.28 & 52.71 & \textcolor{blue}{$-$1.11} \\
f & 6.0 & $-$62.04 & 65.36 & $-$1.05 \\
f & 7.0 & $-$78.11 & 79.32 & $-$1.02 \\
f & 8.0 & $-$96.64 & 92.40 & $-$0.96 \\
f & 9.0 & $-$114.20 & 109.96 & $-$0.96 \\
f & 10.0 & $-$131.85 & 120.31 & $-$0.91 \\
f & 11.0 & $-$147.04 & 138.90 & $-$0.94 \\
f & 12.0 & $-$160.99 & 148.61 & $-$0.92 \\
f & 13.0 & $-$171.84 & 163.52 & $-$0.95 \\
f & 14.0 & $-$183.12 & 171.53 & $-$0.94 \\
f & 15.0 & $-$191.37 & 184.79 & $-$0.97 \\
f & 16.0 & $-$201.32 & 191.85 & $-$0.95 \\
f & 17.0 & $-$209.17 & 201.19 & $-$0.96 \\
f & 18.0 & $-$219.13 & 212.50 & $-$0.97 \\
f & 19.0 & $-$227.26 & 216.62 & $-$0.95 \\
f & 20.0 & $-$237.99 & 231.14 & $-$0.97 \\
p1 & 5.0 & $-$28.49 & 31.97 & \textcolor{blue}{$-$1.12} \\
p1 & 6.0 & $-$34.98 & 37.22 & $-$1.06 \\
p1 & 7.0 & $-$41.51 & 44.68 & $-$1.08 \\
p1 & 8.0 & $-$48.55 & 52.32 & $-$1.08 \\
p1 & 9.0 & $-$55.60 & 59.03 & $-$1.06 \\
p1 & 10.0 & $-$63.31 & 66.50 & $-$1.05 \\
p1 & 11.0 & $-$70.62 & 74.56 & $-$1.06 \\
p1 & 12.0 & $-$78.00 & 81.55 & $-$1.05 \\
p1 & 13.0 & $-$85.17 & 89.56 & $-$1.05 \\
p1 & 14.0 & $-$92.60 & 96.07 & $-$1.04 \\
p1 & 15.0 & $-$99.24 & 105.81 & $-$1.07 \\
p1 & 16.0 & $-$106.20 & 109.87 & $-$1.03 \\
p1 & 17.0 & $-$112.68 & 119.97 & $-$1.06 \\
p1 & 18.0 & $-$118.66 & 122.89 & $-$1.04 \\
p1 & 19.0 & $-$124.64 & 133.35 & $-$1.07 \\
p1 & 20.0 & $-$131.46 & 137.21 & $-$1.04 \\
p2 & 5.0 & $-$15.62 & 16.75 & $-$1.07 \\
p2 & 6.0 & $-$19.79 & 24.20 & \textcolor{orange}{$-$1.22} \\
p2 & 7.0 & $-$23.77 & 26.68 & \textcolor{blue}{$-$1.12} \\
p2 & 8.0 & $-$27.45 & 28.39 & $-$1.03 \\
p2 & 9.0 & $-$30.85 & 34.11 & \textcolor{blue}{$-$1.11} \\
p2 & 10.0 & $-$34.24 & 38.73 & \textcolor{blue}{$-$1.13} \\
p2 & 11.0 & $-$37.97 & 40.67 & $-$1.07 \\
p2 & 12.0 & $-$41.86 & 45.66 & $-$1.09 \\
p2 & 13.0 & $-$45.82 & 50.54 & \textcolor{blue}{$-$1.10} \\
p2 & 14.0 & $-$49.53 & 53.00 & $-$1.07 \\
p2 & 15.0 & $-$53.00 & 57.86 & $-$1.09 \\
p2 & 16.0 & $-$56.85 & 62.32 & $-$1.10 \\
p2 & 17.0 & $-$60.83 & 64.98 & $-$1.07 \\
p2 & 18.0 & $-$64.42 & 70.26 & $-$1.09 \\
p2 & 19.0 & $-$68.05 & 73.36 & $-$1.08 \\
p2 & 20.0 & $-$71.80 & 76.65 & $-$1.07 \\
p3 & 5.0 & $-$8.20 & 8.44 & $-$1.03 \\
p3 & 6.0 & $-$9.95 & 9.48 & $-$0.95 \\
p3 & 7.0 & $-$12.62 & 12.40 & $-$0.98 \\
p3 & 8.0 & $-$15.46 & 16.25 & $-$1.05 \\
p3 & 9.0 & $-$17.63 & 18.57 & $-$1.05 \\
p3 & 10.0 & $-$19.25 & 19.85 & $-$1.03 \\
p3 & 11.0 & $-$20.99 & 21.28 & $-$1.01 \\
p3 & 12.0 & $-$23.03 & 23.04 & $-$1.00 \\
p3 & 13.0 & $-$25.24 & 25.32 & $-$1.00 \\
p3 & 14.0 & $-$27.45 & 27.64 & $-$1.01 \\
p3 & 15.0 & $-$29.49 & 29.40 & $-$1.00 \\
p3 & 16.0 & $-$31.40 & 31.22 & $-$0.99 \\
p3 & 17.0 & $-$33.43 & 33.66 & $-$1.01 \\
p3 & 18.0 & $-$35.52 & 35.46 & $-$1.00 \\
p3 & 19.0 & $-$37.63 & 37.04 & $-$0.98 \\
p3 & 20.0 & $-$39.77 & 38.47 & $-$0.97 \\
p4 & 5.0 & $-$6.05 & 10.25 & \textcolor{red}{$-$1.70} \\
p4 & 6.0 & $-$5.33 & 7.06 & \textcolor{orange}{$-$1.33} \\
p4 & 7.0 & $-$6.02 & 6.10 & $-$1.01 \\
p4 & 8.0 & $-$7.32 & 6.61 & $-$0.90 \\
p4 & 9.0 & $-$8.95 & 7.86 & \textcolor{blue}{$-$0.88} \\
p4 & 10.0 & $-$10.66 & 9.47 & \textcolor{blue}{$-$0.89} \\
p4 & 11.0 & $-$11.99 & 11.33 & $-$0.94 \\
p4 & 12.0 & $-$12.77 & 13.33 & $-$1.04 \\
p4 & 13.0 & $-$13.40 & 14.82 & \textcolor{blue}{$-$1.11} \\
p4 & 14.0 & $-$14.24 & 15.40 & $-$1.08 \\
p4 & 15.0 & $-$15.35 & 15.42 & $-$1.00 \\
p4 & 16.0 & $-$16.63 & 15.62 & $-$0.94 \\
p4 & 17.0 & $-$17.94 & 16.48 & $-$0.92 \\
p4 & 18.0 & $-$19.10 & 18.00 & $-$0.94 \\
p4 & 19.0 & $-$20.05 & 19.86 & $-$0.99 \\
p4 & 20.0 & $-$20.92 & 21.34 & $-$1.02 \\
td1 & 2.5 & $-$39.43 & $-$5.01 & \textcolor{red}{0.13} \\
td1 & 3.4 & $-$50.46 & $-$5.01 & \textcolor{red}{0.10} \\
td1 & 4.2 & $-$55.64 & 25.48 & \textcolor{red}{$-$0.46} \\
td1 & 5.1 & $-$60.84 & 21.82 & \textcolor{red}{$-$0.36} \\
td1 & 6.0 & $-$67.35 & 15.55 & \textcolor{red}{$-$0.23} \\
td2 & 4.2 & $-$50.62 & 21.56 & \textcolor{red}{$-$0.43} \\
td2 & 5.1 & $-$57.73 & 40.57 & \textcolor{orange}{$-$0.70} \\
td2 & 6.0 & $-$61.46 & 35.99 & \textcolor{orange}{$-$0.59} \\
td2 & 6.8 & $-$69.44 & 55.52 & \textcolor{orange}{$-$0.80} \\
td2 & 7.7 & $-$74.02 & 57.45 & \textcolor{orange}{$-$0.78} \\
td3 & 6.0 & $-$63.73 & 38.09 & \textcolor{orange}{$-$0.60} \\
td3 & 7.0 & $-$72.10 & 50.28 & \textcolor{orange}{$-$0.70} \\
td3 & 7.9 & $-$74.42 & 58.24 & \textcolor{orange}{$-$0.78} \\
td3 & 8.9 & $-$79.60 & 64.57 & \textcolor{blue}{$-$0.81} \\
td3 & 9.9 & $-$84.18 & 73.52 & \textcolor{blue}{$-$0.87} \\
td4 & 9.9 & $-$61.41 & 40.69 & \textcolor{orange}{$-$0.66} \\
td4 & 10.8 & $-$61.30 & 46.88 & \textcolor{orange}{$-$0.76} \\
td4 & 11.6 & $-$63.35 & 60.95 & $-$0.96 \\
td4 & 12.4 & $-$64.19 & 60.95 & $-$0.95 \\
td4 & 13.3 & $-$63.68 & 60.03 & $-$0.94 \\
td5 & 13.0 & $-$45.37 & 39.25 & \textcolor{blue}{$-$0.87} \\
td5 & 15.0 & $-$50.37 & 58.11 & \textcolor{blue}{$-$1.15} \\
td5 & 17.0 & $-$48.79 & 54.17 & \textcolor{blue}{$-$1.11} \\
td5 & 18.0 & $-$48.96 & 58.58 & \textcolor{blue}{$-$1.20} \\
td5 & 20.0 & $-$50.48 & 61.70 & \textcolor{orange}{$-$1.22} \\
td6 & 18.0 & $-$70.79 & 84.88 & \textcolor{blue}{$-$1.20} \\
td6 & 19.0 & $-$32.61 & 43.44 & \textcolor{orange}{$-$1.33} \\
td6 & 21.0 & $-$37.74 & 45.92 & \textcolor{orange}{$-$1.22} \\
td6 & 22.0 & $-$53.66 & 70.21 & \textcolor{orange}{$-$1.31} \\
td6 & 24.0 & $-$39.21 & 48.58 & \textcolor{orange}{$-$1.24} \\
td7 & 22.0 & $-$37.92 & 56.06 & \textcolor{orange}{$-$1.48} \\
td7 & 23.0 & $-$32.18 & 41.83 & \textcolor{orange}{$-$1.30} \\
td7 & 25.0 & $-$51.49 & 72.09 & \textcolor{orange}{$-$1.40} \\
td7 & 27.0 & $-$36.45 & 50.51 & \textcolor{orange}{$-$1.39} \\
td7 & 29.0 & $-$39.76 & 57.62 & \textcolor{orange}{$-$1.45} \\
td8 & 26.0 & $-$29.82 & 39.78 & \textcolor{orange}{$-$1.33} \\
td8 & 28.0 & $-$36.02 & 47.63 & \textcolor{orange}{$-$1.32} \\
td8 & 29.0 & $-$44.73 & 65.63 & \textcolor{orange}{$-$1.47} \\
td8 & 31.0 & $-$35.49 & 48.57 & \textcolor{orange}{$-$1.37} \\
td8 & 33.0 & $-$34.48 & 46.36 & \textcolor{orange}{$-$1.34} \\
td9 & 30.0 & $-$31.85 & 47.45 & \textcolor{orange}{$-$1.49} \\
td9 & 32.0 & $-$32.57 & 46.97 & \textcolor{orange}{$-$1.44} \\
td9 & 34.0 & $-$36.22 & 60.67 & \textcolor{red}{$-$1.67} \\
td9 & 35.0 & $-$33.85 & 56.15 & \textcolor{red}{$-$1.66} \\
td9 & 37.0 & $-$32.13 & 47.08 & \textcolor{orange}{$-$1.47} \\
td10 & 35.0 & $-$21.78 & 39.61 & \textcolor{red}{$-$1.82} \\
td10 & 36.0 & $-$22.85 & 42.70 & \textcolor{red}{$-$1.87} \\
td10 & 38.0 & $-$31.20 & 60.62 & \textcolor{red}{$-$1.94} \\
td10 & 39.0 & $-$30.22 & 57.55 & \textcolor{red}{$-$1.90} \\
td10 & 41.0 & $-$24.35 & 46.56 & \textcolor{red}{$-$1.91} \\
td11 & 39.0 & $-$13.80 & 25.39 & \textcolor{red}{$-$1.84} \\
td11 & 40.0 & $-$17.70 & 33.59 & \textcolor{red}{$-$1.90} \\
td11 & 42.0 & $-$29.45 & 64.82 & \textcolor{red}{$-$2.20} \\
td11 & 44.0 & $-$22.76 & 45.85 & \textcolor{red}{$-$2.01} \\
td11 & 46.0 & $-$16.95 & 30.16 & \textcolor{red}{$-$1.78} \\
\hline
\label{tab:allkernels}
\end{longtable}

\end{document}